\documentclass[conference]{IEEEtran}
\IEEEoverridecommandlockouts

\usepackage{cite}
\usepackage{amsmath,amssymb,amsfonts}
\usepackage{graphicx}
\usepackage[table]{xcolor}
\usepackage{booktabs}
\usepackage{multirow}
\usepackage{subcaption}
\usepackage{url}
\usepackage[hidelinks,breaklinks=true]{hyperref}

\begin{document}

\title{How Contrastive Decoding Enhances Large Audio Language Models}

\author{
\IEEEauthorblockN{Tzu-Quan Lin\textsuperscript{1}, Wei-Ping Huang\textsuperscript{1}, Yi-Cheng Lin\textsuperscript{1}, Hung-yi Lee\textsuperscript{1,2}}
\IEEEauthorblockA{\textsuperscript{1} Graduate Institute of Communication Engineering, National Taiwan University, Taiwan\\
\textsuperscript{2} NTU Artificial Intelligence Center of Research Excellence (NTU AI-CoRE), Taiwan\\
tzuquanlin@gmail.com, hungyilee@ntu.edu.tw}
}

\maketitle

\begin{abstract}
While Contrastive Decoding (CD) has been proposed to enhance Large Audio Language Models (LALMs), it has not been evaluated at scale, and the underlying mechanisms driving its success remain unclear. This study systematically evaluates four distinct CD strategies across diverse LALM architectures. We identify Audio-Aware Decoding and Audio Contrastive Decoding as the most effective methods. However, their impact varies significantly across models. To explain this variability, we profile the baseline error composition of each model and measure how readily contrastive decoding corrects each error type. Our analysis demonstrates that CD reliably rectifies errors in which models falsely claim an absence of audio or resort to uncertainty-driven guessing, but is relatively poor at correcting flawed reasoning or confident misassertions. Crucially, CD's benefit closely tracks the composition of a model's baseline error profile: when errors caused by audio ignorance or uncertainty-driven guessing constitute only a small fraction of a model's errors, gains are marginal or even negative. A token-level analysis reveals the underlying mechanism: when the amateur's output is dominated by hesitation markers, CD's suppression naturally targets uncertainty-driven errors while having limited effect on confident misassertions.
\end{abstract}


\section{Introduction}

Audio-text modeling has rapidly evolved from cascaded systems to end-to-end Large Audio Language Models (LALMs)~\cite{gong2023joint, gong2024listen, chu2023qwenaudio, chu2024qwen2audio, tang2024salmonn, xu2025qwen2, lu2025desta, goel2025audio, ghosh2024gama, hu2024wavllm, kong2024audio, abouelenin2025phi4mini, huang2025speechcaps, ghosh2025audioflamingo2, held2024distilling, huang2024dynamic, yang2024building}. By directly mapping continuous audio representations into the embedding space of Large Language Models (LLMs), these architectures enable a richer understanding of paralinguistic features often lost in textual transcriptions. However, LALMs inherit the tendency to hallucinate from their LLM backbones, frequently ignoring audio inputs~\cite{du2024when} or generating plausible but incorrect content~\cite{sahoo2024comprehensive, kuan2024teaching}.
Contrastive Decoding (CD) has emerged as a technique to mitigate these issues by amplifying the difference between an expert and an amateur model through logit subtraction, suppressing negative priors such as generic language biases~\cite{li2023contrastive}.

While Audio-Aware Decoding (AAD)~\cite{hsu2025reducing} has been proposed for LALMs, and various CD strategies exist in text~\cite{li2023contrastive, chuang2024dola, yang2025less} and vision~\cite{leng2024mitigating, wang2024mitigating, zhu2025ibd} domains, their comparative effectiveness remains unexplored in the audio domain, where acoustic signals are continuous and harder to manipulate~\cite{guo2025recent}.
This leads to our first research question: \textit{Which type of contrastive decoding strategy yields the greatest benefit for LALMs?} We evaluate four distinct strategies and find that AAD and Audio Contrastive Decoding (ACD)~\cite{leng2024mitigating} provide the most significant improvements. Furthermore, we observe substantial variability across models: Qwen2.5-Omni~\cite{xu2025qwen2} benefits far more than DeSTA2.5-Audio~\cite{lu2025desta} and Audio Flamingo 3~\cite{goel2025audio}.

This disparity prompts our second question: \textit{What characteristics make a model susceptible to improvement via contrastive decoding?} We profile the baseline error composition of each model and measure how readily contrastive decoding corrects each error type. Our analysis reveals that CD is highly effective when a model's error profile includes a sufficient proportion of audio blindness or uncertainty errors (e.g., Qwen2.5-Omni), but largely ineffective when such errors make up only a small share and the profile is dominated by flawed reasoning or confident misassertions (e.g., DeSTA2.5-Audio, Audio Flamingo 3). A token-level analysis on Qwen2.5-Omni further provides a mechanistic explanation: when the amateur's distribution is dominated by hesitation markers (e.g., \textit{Hmm}, \textit{Sorry}), error types that align with this uncertainty-driven prior are susceptible to logit subtraction, while confident errors are relatively harder to correct.

\begin{figure*}[t]
    \centering
    \includegraphics[width=1.0\linewidth]{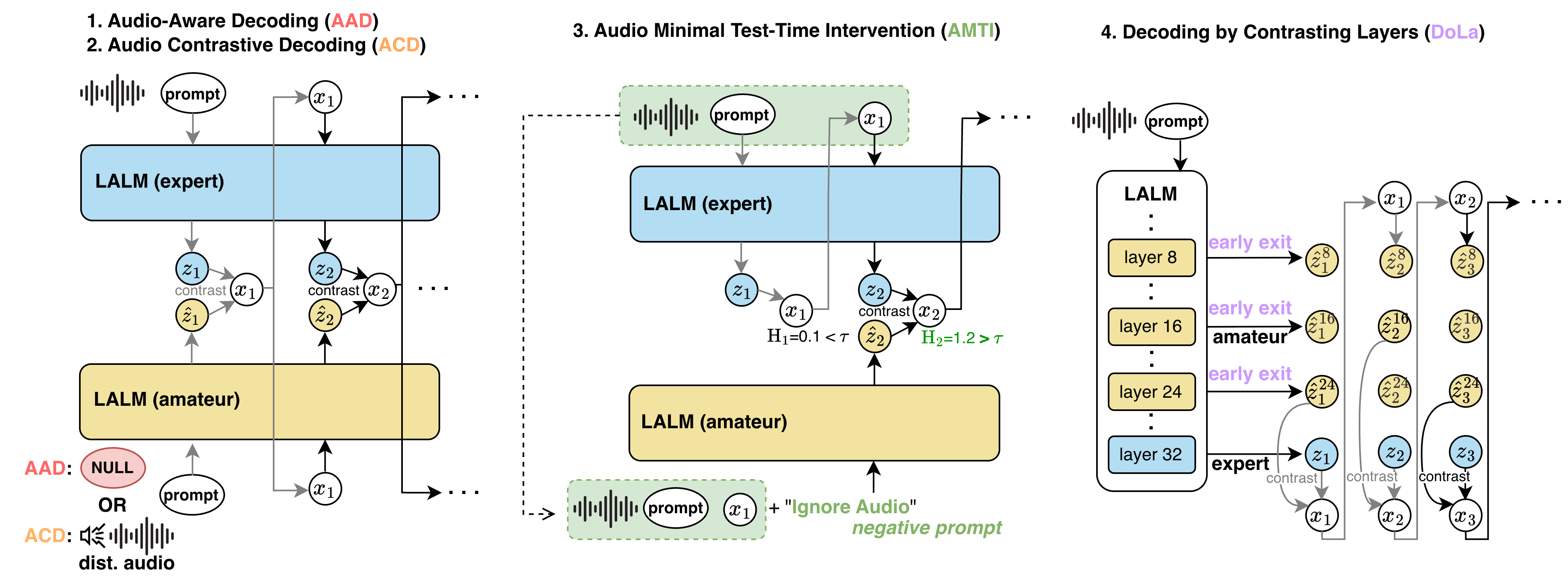}
    \vspace{-1.5em}
    \caption{Overview of different contrastive decoding methods on LALM. In the diagrams, $z_t$ represents the expert logits, while $\hat{z}_t$ denotes the amateur logits.}
    \label{fig:methods}
\end{figure*}

Our contributions are threefold:
\begin{enumerate}
    \item First, we systematically evaluate four CD strategies for multi-token generation in LALMs across diverse architectures and tasks, identifying AAD and ACD as the most effective.
    \item Second, we explain \emph{why} CD selectively improves LALMs. We show that CD's gains closely track a model's baseline error profile: models rich in audio blindness or uncertainty errors benefit significantly, while those dominated by confident misassertions do not. A token-level analysis on Qwen2.5-Omni further reveals the underlying mechanism: when the amateur's output is dominated by hesitation markers, CD predominantly suppresses uncertainty-driven errors while having limited effect on confident misassertions.
    \item Finally, given that AAD and ACD require two forward passes per decoding step, we recommend a lightweight offline assessment of the baseline error profile on a validation subset before committing to deployment.
\end{enumerate}

\section{Background}
\subsection{Large Audio Language Models}

Various Large Audio Language Models (LALMs) with diverse architectures and capabilities have rapidly emerged~\cite{gong2023joint, gong2024listen, chu2023qwenaudio, chu2024qwen2audio, tang2024salmonn, xu2025qwen2, lu2025desta, goel2025audio, ghosh2024gama, hu2024wavllm, kong2024audio, abouelenin2025phi4mini, huang2025speechcaps, ghosh2025audioflamingo2, held2024distilling}. We select three representative state-of-the-art (SOTA) LALMs with significant structural differences: Qwen2.5-Omni~\cite{xu2025qwen2}, which utilizes a ``Thinker'' module for unified reasoning; DeSTA2.5-Audio~\cite{lu2025desta}, which bridges a Whisper encoder and Llama-3.1 via a Q-Former; and Audio Flamingo 3 (AF3)~\cite{goel2025audio}, which integrates a sliding-window Whisper encoder with Qwen2.5-7B. This diversity allows us to analyze how different modeling choices impact the effectiveness of CD. We utilize AF3's standard generation mode (excluding CoT prompt) to ensure fair comparison.

\subsection{Contrastive Decoding Methods}
\label{sec:cd_methods}
Contrastive decoding methods enhance model generation by amplifying the difference between an expert distribution and a flawed amateur distribution~\cite{li2023contrastive}. In the context of LALMs, given audio input $\mathbf{a}$, text instruction $\mathbf{p}$, and previously decoded tokens $\mathbf{x}_1, \dots, \mathbf{x}_t$, we denote the standard LALM logits as the expert $z_{t+1} = \text{LALM}(\mathbf{a}, \mathbf{p}, \mathbf{x}_{\le t})$. We subtract the amateur logits $\hat{z}_{t+1}$ from the expert logits. The final manipulated logits $\tilde{z}_{t+1}$ for generating the next token $x_{t+1}$ are computed as:

\begin{equation}
    \tilde{z}_{t+1} = \alpha \cdot z_{t+1} - \beta \cdot \hat{z}_{t+1}
    \label{eq:cd}
\end{equation}

where $\alpha$ and $\beta$ are hyperparameters controlling the strength of the contrast.
We investigate four distinct strategies to construct the amateur logits $\hat{z}_{t+1}$.

\vspace{0.5em}
\noindent \textbf{Audio-Aware Decoding (AAD).}
Originally proposed by Hsu et al.~\cite{hsu2025reducing} to mitigate object hallucinations in LALMs, AAD assumes hallucinations often stem from the language model's prior rather than the audio content. As shown in the left panel of Figure~\ref{fig:methods}, AAD constructs the amateur by removing the audio modality (setting $\mathbf{a} = \emptyset$), forcing the model to rely solely on the text prompt.

\begin{equation}
    \hat{z}_{t+1} = \text{LALM}(\emptyset, \mathbf{p},\mathbf{x}_{\le t})
\end{equation}

By contrasting this text-only prior with the full multimodal output, AAD highlights audio-grounded information.

\vspace{0.5em}
\noindent \textbf{Audio Contrastive Decoding (ACD).}
Originally proposed as Visual Contrastive Decoding (VCD)~\cite{leng2024mitigating} to mitigate visual hallucinations, this method posits that statistical biases are robust to input noise while correct perception is sensitive to it. As shown in the left panel of Figure~\ref{fig:methods}, we extend this principle to the audio domain (ACD) by contrasting the expert output with an amateur distribution derived from distorted audio inputs $\mathbf{a}_{dist}$.
Specifically, the distorted audio is generated by progressively injecting Gaussian noise into the original signal over 100 steps with a noise intensity scale of 0.1.

\begin{equation}
    \hat{z}_{t+1} = \text{LALM}(\mathbf{a}_{dist}, \mathbf{p},\mathbf{x}_{\le t})
\end{equation}

This method penalizes tokens that remain unchanged despite significant degradation in the audio signal.

\vspace{0.5em}
\noindent \textbf{Audio Minimal Test-Time Intervention (AMTI).}
Derived from the MTI framework~\cite{yang2025less} for LLM reasoning, AMTI intervenes only when necessary for better efficiency.
It monitors the entropy $H_{t+1}$ of the expert distribution. Contrastive decoding is triggered only when the model is uncertain (i.e., $H_{t+1} > \tau$).
Notably, the amateur logits are computed by reusing the KV-cache to ensure efficiency.
As shown in the middle panel of Figure~\ref{fig:methods}, we adapt the negative prompt from MTI by using an instruction $\mathbf{p}^-$ (e.g., ``Ignore Audio'') to induce an amateur that ignores acoustic evidence.

\begin{equation}
    \hat{z}_{t+1} = \begin{cases} \text{LALM}(\mathbf{a}, \mathbf{p}, \mathbf{x}_{\le t},\mathbf{p}^-) & \text{if } H_{t+1} > \tau \\ \phi & \text{otherwise} \end{cases}
\end{equation}

Here $\phi$ denotes no contrast at this step.

\vspace{0.5em}
\noindent \textbf{Decoding by Contrasting Layers (DoLa).}
Originally proposed to enhance LLM factuality~\cite{chuang2024dola}, DoLa exploits the localization of factual knowledge in later transformer layers. Unlike static amateurs, DoLa dynamically selects an intermediate layer to serve as the amateur. Specifically, we obtain the amateur logits $\hat{z}^k_{t+1}$ from the $k$-th layer by projecting its hidden state $\mathbf{h}^k_{t+1}$ into the vocabulary space using the pre-trained language modeling head $\mathbf{W}_{\text{head}}$ of the final layer:

\begin{equation}
    \hat{z}^k_{t+1} = \mathbf{W}_{\text{head}} \cdot \text{Norm}(\mathbf{h}^k_{t+1})
\end{equation}

where $\text{Norm}(\cdot)$ corresponds to the final layer normalization. We then determine the optimal amateur layer $k \in \mathcal{K}$ at each step by maximizing the Jensen-Shannon Divergence (JSD) between the post-softmax expert and candidate probability distributions:

\begin{equation}
     k = \operatorname*{argmax}_{k \in \mathcal{K}} \text{JSD}\!\left(p_{t+1} \,\|\, \hat{p}^k_{t+1}\right), \quad\hat{z}_{t+1} = \hat{z}^k_{t+1}
\end{equation}
where $p_{t+1} = \text{softmax}(z_{t+1})$, $\hat{p}^k_{t+1} = \text{softmax}(\hat{z}^k_{t+1})$, and $\mathcal{K} = \{N/2, N/2 + 2, \dots, N-1\}$ is the candidate layer set for an $N$-layer model. The step size of $+2$ follows the original DoLa paper to optimize efficiency.

This amplifies information processed by deep layers while suppressing superficial patterns from early layers~\cite{pasad2023comparative, lin2024property, lin2025identifying}. The process is shown in the right panel of Figure~\ref{fig:methods}.

\begin{table*}[ht]
    \centering
    \caption{Performance of different contrastive decoding methods. \textbf{Bold} indicates the best performance within each model block. $^{*}$ indicates a statistically significant difference from Greedy (McNemar's test, $p < 0.05$). Columns with gray background indicate average scores.}
    \vspace{-0.5em}
    \label{tab:performance}
    \renewcommand{\arraystretch}{0.8}
    \setlength{\tabcolsep}{4pt}
    \begin{tabular}{llcccccccccc}
        \toprule
        \multirow{2}{*}{\textbf{Model}} & \multirow{2}{*}{\textbf{Method}} & \multicolumn{5}{c}{\textbf{SAKURA}} & \multicolumn{4}{c}{\textbf{MMAU}} & \textbf{MMAR} \\
        \cmidrule(lr){3-7} \cmidrule(lr){8-11} \cmidrule(lr){12-12}
        & & Animal & Emotion & Gender & Language & \cellcolor{gray!20}Avg & Speech & Sound & Music & \cellcolor{gray!20}Avg & \cellcolor{gray!20}Avg \\
        \midrule

        \multirow{5}{*}{Qwen2.5-Omni-7B}
          & Greedy & 95.0 & 22.6 & 67.2 & 93.6 & \cellcolor{gray!20}69.6 & 72.4 & 73.3 & 57.8 & \cellcolor{gray!20}67.8 & \cellcolor{gray!20}45.1 \\
          & AAD    & 95.6 & \textbf{52.0}$^{*}$ & \textbf{88.0}$^{*}$ & \textbf{95.0} & \cellcolor{gray!20}\textbf{82.7}$^{*}$ & 71.2 & \textbf{77.5} & \textbf{66.2}$^{*}$ & \cellcolor{gray!20}\textbf{71.6}$^{*}$ & \cellcolor{gray!20}\textbf{54.3}$^{*}$ \\
          & ACD    & \textbf{96.4}$^{*}$ & 34.2$^{*}$ & 84.8$^{*}$ & \textbf{95.0} & \cellcolor{gray!20}77.6$^{*}$ & \textbf{73.9} & 74.5 & 62.9$^{*}$ & \cellcolor{gray!20}70.4$^{*}$ & \cellcolor{gray!20}53.1$^{*}$ \\
          & AMTI   & 93.0$^{*}$ & 38.8$^{*}$ & 81.8$^{*}$ & 92.8 & \cellcolor{gray!20}76.6$^{*}$ & 64.0$^{*}$ & 70.3 & 59.6 & \cellcolor{gray!20}64.6$^{*}$ & \cellcolor{gray!20}39.8$^{*}$ \\
          & DoLa   & 94.4 & 31.0$^{*}$ & 86.2$^{*}$ & 93.8 & \cellcolor{gray!20}76.4$^{*}$ & 68.8 & 77.2 & 61.4 & \cellcolor{gray!20}69.1 & \cellcolor{gray!20}51.6$^{*}$ \\
        \midrule

        \multirow{5}{*}{DeSTA2.5-Audio}
          & Greedy & 60.6 & 55.4 & 87.6 & \textbf{96.2} & \cellcolor{gray!20}75.0 & 66.7 & 67.0 & 52.7 & \cellcolor{gray!20}62.1 & \cellcolor{gray!20}48.1 \\
          & AAD    & 64.4$^{*}$ & 54.0 & 90.6$^{*}$ & 95.4 & \cellcolor{gray!20}76.1 & 64.0 & 63.1 & 50.6 & \cellcolor{gray!20}59.2$^{*}$ & \cellcolor{gray!20}46.2 \\
          & ACD    & \textbf{65.6}$^{*}$ & \textbf{59.2}$^{*}$ & \textbf{91.4}$^{*}$ & \textbf{96.2} & \cellcolor{gray!20}\textbf{78.1}$^{*}$ & \textbf{67.9} & \textbf{67.3} & \textbf{57.2} & \cellcolor{gray!20}\textbf{64.1} & \cellcolor{gray!20}\textbf{49.0} \\
          & AMTI   & 54.4$^{*}$ & 57.8 & 87.4 & 95.0 & \cellcolor{gray!20}73.7 & 67.3 & 66.7 & 55.1 & \cellcolor{gray!20}63.0 & \cellcolor{gray!20}48.1 \\
          & DoLa   & 59.2 & 56.0 & 72.6$^{*}$ & 95.8 & \cellcolor{gray!20}70.9$^{*}$ & 66.4 & 64.3 & 48.5 & \cellcolor{gray!20}59.7 & \cellcolor{gray!20}47.2 \\
        \midrule

        \multirow{5}{*}{Audio Flamingo 3}
          & Greedy & 84.4 & 61.0 & 58.2 & 93.2 & \cellcolor{gray!20}74.2 & \textbf{65.8} & \textbf{83.5} & \textbf{75.8} & \cellcolor{gray!20}\textbf{75.0} & \cellcolor{gray!20}60.8 \\
          & AAD    & 86.8 & \textbf{63.0} & \textbf{62.0} & 95.4$^{*}$ & \cellcolor{gray!20}\textbf{76.8}$^{*}$ & \textbf{65.8} & \textbf{83.5} & 72.5$^{*}$ & \cellcolor{gray!20}73.9 & \cellcolor{gray!20}\textbf{61.0} \\
          & ACD    & \textbf{87.6}$^{*}$ & 62.0 & 59.4 & \textbf{96.2}$^{*}$ & \cellcolor{gray!20}76.3$^{*}$ & 65.2 & 82.9 & 70.4$^{*}$ & \cellcolor{gray!20}72.8$^{*}$ & \cellcolor{gray!20}60.0 \\
          & AMTI   & 79.2$^{*}$ & 59.6 & 59.4 & 93.8 & \cellcolor{gray!20}73.0$^{*}$ & 61.6 & 83.2 & 71.6 & \cellcolor{gray!20}72.1 & \cellcolor{gray!20}58.3 \\
          & DoLa   & 67.0$^{*}$ & 52.0$^{*}$ & 57.4 & 93.2 & \cellcolor{gray!20}67.4$^{*}$ & 59.8$^{*}$ & 82.0 & 70.7$^{*}$ & \cellcolor{gray!20}70.8$^{*}$ & \cellcolor{gray!20}54.9$^{*}$ \\
        \bottomrule
    \end{tabular}
\end{table*}

\section{Analysis Methods}
Evaluating CD's internal processing is challenging, as accuracy metrics fail to elucidate the underlying mechanisms of error correction. To identify which failure modes, such as audio blindness or flawed reasoning, are more easily corrected into accurate responses, we introduce an analysis framework that profiles the baseline composition of each error state and measures its Correction Rate under CD, derived from LLM-as-a-Judge evaluations.

\subsection{Categorization of Response States}
To analyze the impact of contrastive decoding on model behavior, we categorize responses into distinct states based on error type. We enforce a strict priority order ($1\rightarrow4$) to ensure mutually exclusive classification:

\begin{enumerate}
    \item Hallucinated No Audio ($W_{NoAudio}$): The model falsely claims no audio was provided or asks to play the sound, excluding cases where it acknowledges the audio but finds it unclear. \textit{(e.g., ``I can't hear the audio.'')}
    \item Reasoning but Wrong ($W_{Reason}$): The model supports a wrong answer with specific evidence, excluding circular reasoning or simple intuition (see $W_{Direct}$ below). \textit{(e.g., ``Water flowing around rocks is characteristic of a river, so B.'')}
    \item Direct Assertive Wrong ($W_{Direct}$): The model asserts a wrong answer without specific evidence, including short assertions, circular reasoning, and simple intuition. \textit{(e.g., ``The answer is C.'' / ``It is A, so the answer is A.'')}
    \item Guessing or Refusal ($W_{Guess}$): The model explicitly states it is ``not sure,'' ``guessing,'' or refuses to answer. This also includes cases where the model attempts to reason but concludes by refusing to answer due to insufficient information. \textit{(e.g., ``I'm not sure, but I'd guess C.'')}
    \item Correct: This state encompasses all responses where the model successfully answers the question.
\end{enumerate}

Ultimately, this 1-to-4 priority order ensures a collectively exhaustive classification of all incorrect responses. It divides errors into four logical buckets: (1) unperceived audio, (2) perceived audio with a reasoned incorrect assertion, (3) perceived audio with an unreasoned incorrect assertion, and (4) perceived audio with an inability to provide a definitive answer.

\subsection{Automated Evaluation via LLM-as-a-Judge}
Manually classifying thousands of responses is infeasible. We employ gpt-4o-2024-11-20 with greedy decoding as an automated judge~\cite{chiang2023large, chiang2023closer}.
The judge is provided with the LALM's response, the question, the correctness status (Correct/Wrong), and strict definitions for the states above.

The judge selects the single most suitable category based on the correctness status and specific linguistic markers. For instance, it distinguishes between $W_{Reason}$ and $W_{Guess}$ by checking for the presence of logical justification versus simple expressions of uncertainty.

To validate our automated judge, six human experts independently annotated 300 randomly sampled responses. On Qwen2.5-Omni, the Human–GPT Kappa is \textbf{0.707} in aggregate and \textbf{0.62}--\textbf{0.87} per state, matching the inter-human values (0.712, 0.66--0.80). For DeSTA2.5-Audio and Audio Flamingo 3, whose error profiles are highly skewed, Kappa is uninformative; we therefore report raw agreement of \textbf{85.0\%} and \textbf{93.5\%}, respectively.

\subsection{Error Profile and Correction Rate}
\label{sec:error_profile}
Our analysis characterizes how CD reshapes model behavior relative to greedy decoding along two axes: (1) the composition of the baseline error profile, and (2) how readily each error state is corrected.

For each error state $i \in \{W_{NoAudio},\allowbreak W_{Reason},\allowbreak W_{Direct},\allowbreak W_{Guess}\}$, we compute two quantities over the samples that greedy decoding answers incorrectly:

\begin{itemize}
    \item \textbf{Baseline error-state proportion} $\pi_i$: the percentage of all initially wrong samples that fall into state $i$.
    \item \textbf{Correction Rate} $\text{CR}_i$: the percentage of state $i$'s samples that CD reclassifies as Correct.
\end{itemize}

We omit $\text{CR}_i$ for any state with $\pi_i < 5\%$, as its rate is too noisy to be meaningful at that sample size.

Separately, for samples that greedy decoding already answers correctly, we report the \textit{Retention Rate}—the percentage of these samples that remain Correct under CD (Section~\ref{sec:correct_retention}).

\section{Experimental Setup}
\subsection{Benchmarks}

We utilize three benchmarks spanning from perception to reasoning: SAKURA~\cite{yang2025sakura} evaluates perceptual capabilities across four tracks (Animal, Emotion, Gender, Language; 500 samples each). MMAU~\cite{sakshi2025mmau} (1,000 samples) tests advanced cognitive abilities requiring complex information extraction and reasoning, while MMAR~\cite{ma2025mmar} (1,000 samples) challenges deep reasoning in mixed-modality scenarios (e.g., speech overlapping with music). For evaluation, the LALM generates a response for each sample. Following the original SAKURA framework~\cite{yang2025sakura}, although all benchmarks are multiple-choice, we use an LLM judge rather than exact-match parsing to allow flexibility in output format: the response, alongside the original question and ground truth, is fed to gpt-4o-2024-11-20 to determine binary correctness. We verified that exact-match parsing preserves all trends reported below.

\begin{figure*}[t]
    \centering
    \includegraphics[width=0.62\linewidth]{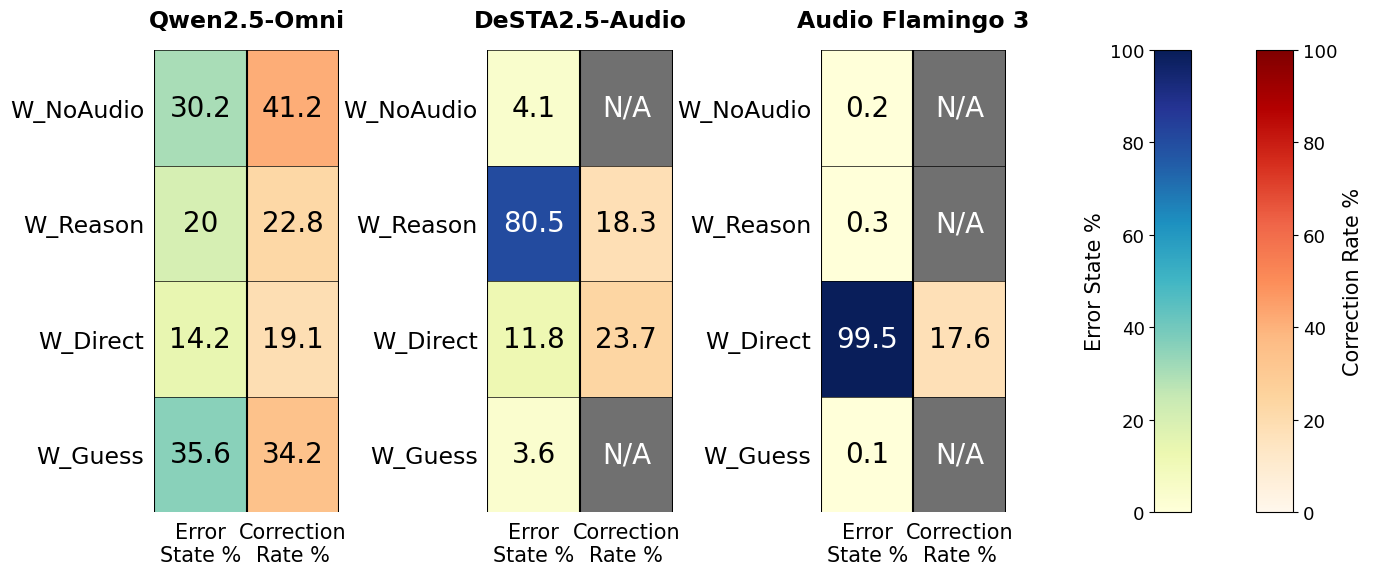}
    \vspace{-0.5em}
    \caption{Baseline error-state proportion ($\pi_i$) and Correction Rate ($\text{CR}_i$), restricted to initially wrong samples: per-model breakdown averaged across all tasks. $\text{CR}_i$ is marked N/A for states with $\pi_i < 5\%$, which are treated as noise.}
    \label{fig:transition_models}
\end{figure*}

\subsection{Implementation Details}
We implemented contrastive decoding methods on Qwen2.5-Omni-7B, DeSTA2.5-Audio, and Audio Flamingo 3 by adhering closely to their original frameworks. Our code is available at \url{https://github.com/nervjack2/LALM-Contrastive-Decoding-Error-Profiles}.
Following prior work~\cite{hsu2025reducing, leng2024mitigating, chuang2024dola}, we set $\alpha=2.0$ and $\beta=1.0$ for AAD, ACD, and AMTI, and $\alpha=1.0$ and $\beta=1.0$ for DoLa, fixing these values globally across all architectures and tasks to ensure a consistent and fair comparison across methods.
For AMTI, the entropy threshold $\tau$ was set to 1.0.
Adhering to the original configurations of VCD and DoLa, we incorporate the Adaptive Plausibility Constraint~\cite{li2023contrastive} into both ACD and our audio implementation of DoLa to preserve each method's intended design.
Our baseline is greedy decoding: the LALM emits the highest-probability token at each step, with no sampling. This isolates the net impact of logit subtraction, avoiding stochastic sampling variance that would otherwise confound CD's deterministic improvements.
\section{Findings}
\subsection{Performance}

Table~\ref{tab:performance} addresses our first research question: \textit{Which type of contrastive decoding strategy yields the greatest benefit for LALMs?} Statistical significance against the greedy baseline is assessed via McNemar's test~\cite{mcnemar1947note} ($p < 0.05$) and marked with $^{*}$. We find that AAD and ACD generally provide the most significant improvements, indicating that contrastive methods directly manipulating the audio input are typically more effective for LALMs. AMTI and DoLa, by contrast, exhibit inconsistent performance and occasionally degrade results relative to the greedy baseline.

However, the effectiveness of AAD and ACD depends heavily on the model. Qwen2.5-Omni achieves substantial gains across most tasks in Table~\ref{tab:performance}: its SAKURA average rises from $69.6\%$ (Greedy) to $82.7\%$ under AAD, and its MMAR average from $45.1\%$ to $54.3\%$. DeSTA2.5-Audio and Audio Flamingo 3, by contrast, show much more limited responsiveness: DeSTA2.5-Audio's SAKURA average rises by only $+3.1$ points ($75.0\% \rightarrow 78.1\%$) under ACD and just $+1.1$ points ($75.0\% \rightarrow 76.1\%$) under AAD, and Audio Flamingo 3's MMAU average actually declines under both AAD ($73.9\%$) and ACD ($72.8\%$) relative to its $75.0\%$ Greedy baseline. Across Table~\ref{tab:performance}, the widest $95\%$ confidence interval is $\pm3.1\%$: Qwen2.5-Omni's SAKURA average under AAD ($82.7\pm1.5$) is cleanly separated from its Greedy baseline ($69.6\pm1.6$), whereas the DeSTA2.5-Audio and Audio Flamingo 3 intervals overlap. This disparity prompts our second question: \textit{What characteristics make a model susceptible to improvement via contrastive decoding?} We investigate the underlying causes in the following section.

\subsection{Error Profile and Correction Rate Analysis}
\label{sec:transition_analysis}

To understand what characteristics make an LALM susceptible to improvement via CD, we analyze the baseline error-state proportion $\pi_i$ and Correction Rate $\text{CR}_i$ (Section~\ref{sec:error_profile}) of each error state, averaged across AAD and ACD (the two most effective methods) and restricted to initially wrong samples; initially correct samples are analyzed separately in Section~\ref{sec:correct_retention}.

Figure~\ref{fig:transition_models} presents the per-model view averaged across six tasks. Qwen2.5-Omni exhibits a highly distributed error pattern, with all four states comprising a non-trivial share of baseline errors ($\pi_i$ between 14.2\% and 35.6\%); $W_{NoAudio}$ and $W_{Guess}$ are the most frequently corrected states, transitioning to Correct at high rates ($\text{CR}=41.2\%$ and $34.2\%$), whereas $W_{Reason}$ and $W_{Direct}$ prove more resistant ($\text{CR}=22.8\%$ and $19.1\%$). DeSTA2.5-Audio errors are heavily concentrated in the $W_{Reason}$ state ($\pi=80.5\%$, $\text{CR}=18.3\%$), and Audio Flamingo 3 funnels almost all errors into $W_{Direct}$ assertions ($\pi=99.5\%$, $\text{CR}=17.6\%$). Both models produce too few $W_{NoAudio}$ and $W_{Guess}$ errors ($\pi<5\%$ each) to report a reliable Correction Rate there; however, their $W_{Reason}$/$W_{Direct}$ rates fall in a tight $17.6$--$23.7\%$ band that closely matches Qwen's own $W_{Reason}$/$W_{Direct}$ rates, suggesting this resistance is an intrinsic property of the error type rather than a model-specific artifact. This shared resistance, well below Qwen's $W_{NoAudio}$/$W_{Guess}$ correction rates, explains the limited responsiveness of DeSTA2.5-Audio and Audio Flamingo 3 compared to Qwen.

These observations lead to our central conclusion: \textbf{CD's effectiveness closely tracks the baseline error profile of the model.} Qwen2.5-Omni's distributed profile, rich in correctable $W_{NoAudio}$ and $W_{Guess}$ errors, explains its broad responsiveness to CD; DeSTA2.5-Audio and Audio Flamingo 3's profiles, dominated by resistant $W_{Reason}$/$W_{Direct}$ errors, explain their limited responsiveness. This finding has a direct practical implication: since methods like AAD and ACD require two forward passes per decoding step, blindly applying CD is wasteful and potentially harmful~\cite{chen2026caad}. We therefore recommend lightweight offline profiling on a small validation subset to gauge whether a model's error profile is CD-amenable before committing to the additional cost.

\subsection{A Closer Look at Qwen's Largest Gains}
\label{sec:case_study}

The model-level view above aggregates across six tasks, which could still mask task-to-task variation within a single model. To examine the same mechanism at its most extreme, we zoom into the two model-task pairs with the single largest CD gains in Table~\ref{tab:performance}—both belonging to Qwen2.5-Omni: SAKURA-Emotion (Greedy $22.6\% \rightarrow$ AAD $52.0\%$, $+29.4$ points) and SAKURA-Gender (Greedy $67.2\% \rightarrow$ AAD $88.0\%$, $+20.8$ points).

\begin{figure}[t]
    \centering
    \includegraphics[width=1.0\linewidth]{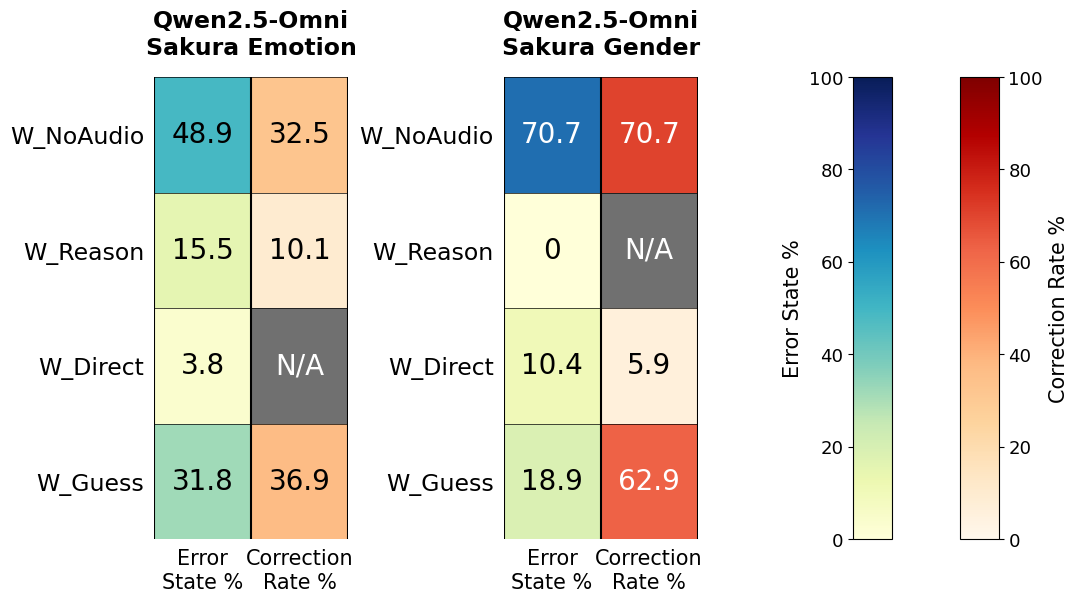}
    \vspace{-0.5em}
    \caption{Case study: baseline error-state proportion ($\pi_i$) and Correction Rate ($\text{CR}_i$), restricted to initially wrong samples, for the two model-task pairs with the largest CD gain (Qwen2.5-Omni on SAKURA-Emotion and SAKURA-Gender) in Table~\ref{tab:performance}. States comprising less than 5\% of wrong samples are treated as noise and their Correction Rate is marked N/A.}
    \label{fig:case_study}
\end{figure}

On both tasks (Figure~\ref{fig:case_study}), Qwen's baseline errors are overwhelmingly $W_{NoAudio}$ and $W_{Guess}$: together they account for $80.7\%$ of errors on Emotion ($\pi=48.9\%$ and $31.8\%$) and $89.6\%$ on Gender ($\pi=70.7\%$ and $18.9\%$), both well above Qwen's own $65.8\%$ model-level average ($\pi=30.2\%$ and $35.6\%$ combined, Figure~\ref{fig:transition_models}), and both states are corrected at high rates on each task ($\text{CR}=32.5\%$/$36.9\%$ on Emotion; $70.7\%$/$62.9\%$ on Gender). The trend is thus even more pronounced at this extreme, consistent with these two tasks delivering Qwen's largest gains.

\subsection{Why Does CD Selectively Correct Errors?}
\label{sec:mechanism}

\begin{table}[t]
\centering
\caption{First-token change analysis under AAD (Qwen2.5-Omni), averaged across all tasks.
$p_1$ and $\hat{p}_1$ denote the expert and amateur output distributions at position 1, respectively; $x_1 = \operatorname{argmax}\,p_1$ is the first token predicted by the expert.
$\mathbb{E}[\hat{p}_1(x_1)]$ is the amateur's probability assigned to $x_1$, averaged over samples.
Chg\% is the rate at which $x_1$ changes after CD.
$\text{W}{\to}\text{C}_{\text{chg}}$ and $\text{W}{\to}\text{C}_{\text{unch}}$ are the correction rates conditional on the first token being changed or unchanged, respectively.}
\label{tab:first_token_change}
\setlength{\tabcolsep}{4pt}
\begin{tabular}{lcccc}
\toprule
State & $\mathbb{E}[\hat{p}_1(x_1)]$ & Chg\% & $\text{W}{\to}\text{C}_{\text{chg}}$ & $\text{W}{\to}\text{C}_{\text{unch}}$ \\
\midrule
$W_{NoAudio}$  & 0.6561 & 91.5 & 56.0 & 10.5 \\
$W_{Guess}$    & 0.6100 & 86.5 & 43.3 & 22.5 \\
$W_{Reason}$   & 0.4019 & 82.0 & 23.6 & 22.6 \\
$W_{Direct}$   & 0.1557 & 56.2 & 24.6 & 14.1 \\
\bottomrule
\end{tabular}
\end{table}

To understand \textit{why} CD corrects certain error states but not others, we focus this token-level analysis on Qwen2.5-Omni, whose evenly distributed baseline error profile (Section~\ref{sec:transition_analysis}) lets us contrast correctable ($W_{NoAudio}$, $W_{Guess}$) against resistant ($W_{Reason}$, $W_{Direct}$) states within a single model. 

We first inspect the amateur's top-10 predicted tokens under AAD (Qwen2.5-Omni). Regardless of error type, the amateur consistently elevates hesitation and uncertainty markers, \textit{Hmm} ($>$80\%) and \textit{Sorry} ($>$20\%), while the remaining slots are occupied by generic sentence starters (e.g., \textit{I}, \textit{Well}, \textit{You}, \textit{It}, \textit{The}). The same inspection under ACD yields an even stronger hesitation prior (\textit{Hmm} 99.8\%, \textit{Sorry} 39.7\%), indicating this pattern is not specific to AAD. Since CD suppresses the amateur's output, it predominantly targets these uncertain openings, leading us to hypothesize that this hesitation-driven prior is the underlying mechanism behind the high correction rates observed for $W_{NoAudio}$ and $W_{Guess}$.

To verify this hypothesis, we conduct a first-token analysis focusing on $x_1 = \operatorname{argmax}\, p_1$, the first token predicted by the expert, where $p_1 = \mathrm{softmax}(z_1)$ and $\hat{p}_1 = \mathrm{softmax}(\hat{z}_1)$ denote the expert and amateur output distributions at position 1, respectively. Results are reported in Table~\ref{tab:first_token_change}. $W_{NoAudio}$ and $W_{Guess}$ exhibit markedly elevated $\mathbb{E}[\hat{p}_1(x_1)]$ (0.6561 and 0.6100), confirming that the amateur's prior strongly aligns with their erroneous first tokens; $W_{Direct}$, by contrast, shows little alignment (0.1557). Accordingly, $W_{NoAudio}$ and $W_{Guess}$ record the highest first-token change rates (91.5\% and 86.5\%), while $W_{Direct}$ is the lowest (56.2\%), confirming that CD more frequently redirects the opening token in correctable states.

The correction rates conditioned on first-token change provide the most decisive evidence. For $W_{NoAudio}$, the rate jumps from 10.5\% (unchanged) to 56.0\% (changed)—a 45.5 percentage-point gap; $W_{Guess}$ shows a similar pattern (22.5\% unchanged vs.\ 43.3\% changed). Yet for $W_{Reason}$, despite a high first-token change rate (82\%), the correction rate is virtually identical whether or not the first token changes (22.6\% unchanged vs.\ 23.6\% changed). We attribute this contrast to the expert's opening token: $W_{NoAudio}$ and $W_{Guess}$ are by nature hesitation-driven, so suppressing the amateur's hesitation marker directly targets their failure mode. $W_{Reason}$ instead opens with a generic starter unrelated to the flawed reasoning that follows, so swapping one generic opener for another offers little leverage.

\subsection{Retention of Initially Correct Predictions}
\label{sec:correct_retention}

The analysis above (Section~\ref{sec:transition_analysis}) is restricted to initially wrong samples. Separately, we measure the Retention Rate of initially correct samples under CD. All three models retain the vast majority of their correct predictions: Qwen2.5-Omni, DeSTA2.5-Audio, and Audio Flamingo 3 remain correct in 94.4\%, 91.1\%, and 93.8\% of cases, respectively, confirming that regressions are limited. Most regressions concentrate in a single error state per model: DeSTA2.5-Audio regresses mainly into $W_{Reason}$ (7.2\% of its initially correct samples), while Audio Flamingo 3 regresses mainly into $W_{Direct}$ (6.1\%), mirroring each model's dominant baseline error state (Section~\ref{sec:transition_analysis}). We hypothesize these limited regressions occur when CD inadvertently over-penalizes functionally necessary generic tokens, occasionally forcing the expert into $W_{Reason}$ or $W_{Direct}$.

\section{Conclusion}
This work systematically evaluates four CD strategies for LALMs, identifying AAD and ACD as the most effective. By profiling the baseline error-state proportion and Correction Rate, we show that CD reliably corrects errors rooted in audio ignorance ($W_{NoAudio}$) or uncertainty-driven guessing ($W_{Guess}$), but is relatively poor at correcting flawed reasoning ($W_{Reason}$) or confident misassertions ($W_{Direct}$). Consequently, models whose correctable errors constitute only a small fraction of the error profile see marginal or negative returns. A token-level analysis on Qwen2.5-Omni traces this selectivity to the amateur's hesitation prior (e.g., \textit{Hmm}, \textit{Sorry}): correctable errors align with these markers, while other error types do not. Since AAD and ACD require two forward passes per decoding step, we recommend profiling the error distribution on a small validation subset offline before deployment.

\section{Limitations}
Our token-level mechanistic analysis (Section~\ref{sec:mechanism}) uses Qwen2.5-Omni as a case study, since DeSTA2.5-Audio and Audio Flamingo 3 have too few $W_{NoAudio}$/$W_{Guess}$ samples ($\pi<5\%$, Section~\ref{sec:transition_analysis}) for a statistically meaningful analysis; future work could test other models with a similarly distributed error profile to verify generality. Our benchmarks are also multiple-choice tasks; extending this analysis to open-ended generation is left for future work.

\section*{Acknowledgment}
Tzu-Quan Lin is supported by a joint Ph.D. scholarship from NTU GICE and Delta Electronics, Inc.
This work was supported by the Ministry of Education (MOE) of Taiwan under the project Taiwan Centers of Excellence in Artificial Intelligence, through the NTU Artificial Intelligence Center of Research Excellence (NTU AI-CoRE).
The authors thank Tzu-Chieh Wei, Chun-Wei Chen, Leonardo Haw-Yang Foo, Ming-Douo Tchouang, Wenze Ren, and Lok-Lam Ieong for their valuable assistance and insightful discussions.
Generative AI tools were employed solely for language polishing and to enhance the clarity and readability of the manuscript.
The authors remain fully accountable for the study design, experimental methodology, data analysis, and the conclusions presented.
No AI-based tools were involved in developing or contributing to the manuscript's substantive scientific content.

\bibliographystyle{IEEEtran}
\bibliography{mybib}

\end{document}